\documentclass[twocolumn,preprintnumbers,amsmath,amssymb]{revtex4}
\usepackage{graphicx}
\usepackage{dcolumn}
\usepackage{bm}

\begin{document}
\title{The influence of polaron size on the conductivity of poly-DNA}
\author{Julia A. Berashevich, Adam D. Bookatz, and Tapash Chakraborty$^\ddag$}
\affiliation{Department of Physics and Astronomy, University of
Manitoba, Winnipeg, MB R3T 2N2, Canada}

\begin{abstract}
The velocity of polaron migration in the long poly-DNA chain 
($\sim$40 base pairs) in an applied electric field has been studied  
within a polaron model. We found that the polaron velocity strongly 
depends on the polaron size. A small polaron shows a slow propagation 
and strong tolerance to the electric field, while a large polaron 
is much faster and less stable with increasing electric field. 
Moreover, the conductance of the DNA molecule within the polaron model 
is found to be sensitive to structural disorders in the DNA geometry, 
but that dependence diminishes with increasing temperature.
\end{abstract}
\maketitle

The application of the biological systems in developing nanoelectronic 
devices has been recognized as one of the most intriguing and promising 
techniques in recent years. This is due to the molecular recognition 
and self-assembly properties that allows them to perform reparation of 
the damaged structures by the non-invasive technology and self-incorporation 
of molecular blocks into the well-structured systems. The DNA molecule 
is one such biological system. The interest on DNA, while originally 
derived as it is the source of genetic information, has been significantly 
elevated due to the discovery of DNA conductance \cite{review,jortner,%
berlin,yoo,kasumov,kawai,porath,conw,bishop1}. 

Understanding the mechanism of charge migration in a complex molecule 
such as DNA is a difficult problem. Just as in the field of condensed 
matter, in simple DNA sequences, if the donor and the acceptor are 
separated by a single potential barrier, the charge transfer has been 
adequately explained by the competition of quantum tunneling and 
incoherent hopping \cite{jortner,berlin} that has also been experimentally 
confirmed \cite{lewis, giese}. However, the observed conductance of 
long DNA molecule with both poly- and mixed sequences, varies from 
insulator to a metallic behavior \cite{yoo, kasumov,kawai,porath}. 
Further, the results for the decrease of conductance with decreasing 
temperature has been somewhat contradictory \cite{yoo, porath}. For the 
description of transport phenomena in polymer chains, a polaron model 
has been known to be very successful \cite{su,Ono}. For DNA, 
this model has shown recently to provide promising results as well, 
especially for explaining the temperature dependence of the DNA conductance 
\cite{yoo,conw,bishop1}. Moreover, additional advantage of the polaron 
model, in comparison to the popular models such as the tight-binding 
approach or the system of kinetic equations, is that it includes the 
interaction of the migrating charge with the DNA lattice.  

In this paper, we have used the polaron model for simulation of the 
dynamics of hole propagation in the dry poly-DNA molecules. We have 
found that the most probable reasons for variation of the DNA conductance 
in different experiments \cite{yoo, kasumov,kawai,porath} can be due 
to, (i) different external and internal conditions that can change
formations of large polarons to smaller ones, thereby decreasing the 
conductance. Therefore, the poly(dG)-poly(dC) and poly(dA)-poly(dT) 
chains due to different parameters for charge propagation should show 
different conductance behavior, as indicated in the experiment \cite{yoo};
(ii) Even a small disorder in the DNA structure can trap the polaron, 
again resulting in a decrease of the conductance. However, a disorder-free 
DNA structure shows a constant velocity of polaron propagation 
in a constant electric field, which gradually increases with 
increasing electric field.

We have simulated the polaron propagation through DNA using the 
Peyrard-Bishop-Holstein (PBH) model, which combines a quantum-mechanical 
treatment of charge motion with a classical treatment of the lattice 
distortion dynamics. Application of this model, which describes the 
charge tunneling from one DNA base site to another, is however limited 
to low temperatures. To avoid the temperature restriction we have investigated 
the poly-DNA molecule. In a poly-chain, each strand contains only one base 
type. Therefore, due to the large difference of ionization potential 
between the purine and pyrimidine bases ($\geqslant$ 1.0 eV \cite{my_paper1}) 
belonging to the opposite strands, the main mechanism of charge migration 
at low and at room temperature is  longitudinal one-dimensional tunneling 
along a single strand containing purine bases. 

An electric field is applied to the system using the method of Ref.~\cite{Ono}, 
specifically, the charge transfer integral $V$ is multiplied by a complex 
exponential phase factor that includes the electromagnetic 
vector potential $A$. For a uniform constant electric field $E$, 
we take $A(t) = -cEt$, thereby satisfying $E~=~-{dA}/c{dt}$. 
Here, $c$ is the speed of light, and we have taken $A$=0 at time $t$=0. 
The uniform electric field is included via its uniform vector potential, 
rather than with its non-uniform scalar potential \cite{bishop1}, 
in order that periodic boundary conditions can be utilized. 

With the modifications for the electric field, the coupled system 
of non-linear equations based on the PBH model is
\begin{eqnarray}
i\hbar\frac{d\Psi_n}{dt} &=& -Ve^{-i({ed}/{\hbar c})A(t)}
\Psi_{n-1} - Ve^{i({ed}/{\hbar c})A(t)}\Psi_{n+1} 
\nonumber\\&+& \chi y_n\Psi_n + \epsilon_n\Psi_n,
\label{eq:schrod}
\end{eqnarray}
and
\begin{eqnarray}
m\frac{d^2y_n}{dt^2} &=& -\frac{dV_M(y_n)}{dy_n}-
\frac{dW(y_n,y_{n-1})}{dy_n}-\frac{dW(y_{n+1},y_n)}{dy_n} 
\nonumber\\ &-& \chi |\Psi_n|^2 - m\gamma\frac{dy_n}{dt}
\label{eq:newton}
\end{eqnarray}
where $\Psi_n$ is the probability amplitude for the charge 
on the $n$-th base pair, $V$ is the nearest-neighbor transfer 
integral between the base pairs, $\chi$ is the charge-vibrational 
coupling constant, $\epsilon_n$ is the on-site energy for base 
site $n$, $y_n$ is the amount by which the $n$-th base site is 
displaced from its equilibrium position, $m$ is the polaron mass on the single site, 
$\gamma$ is the friction parameter, $V_M(y_n)$ is the Morse potential, 
and $W(y_n,y_{n-1})$ is the interaction of neighboring stacked base-pairs, 
$e$ is the electronic charge, and $d$ is the interbase spacing ($d$=3.4\AA). 
The expressions and parameters for $V_M(y_n)$  and $W(y_n,y_{i-n})$ are 
taken from Ref.~\cite{bishop1}, and throughout this paper we take 
$\gamma=1$ ps$^{-1}$. In poly-DNA, $\chi$ and $V$ are constant with respect to 
the lattice site, and in the absence of disorder, we choose $\epsilon_n$ 
to be at the origin.

\begin{figure}
\includegraphics[width=0.44\textwidth]{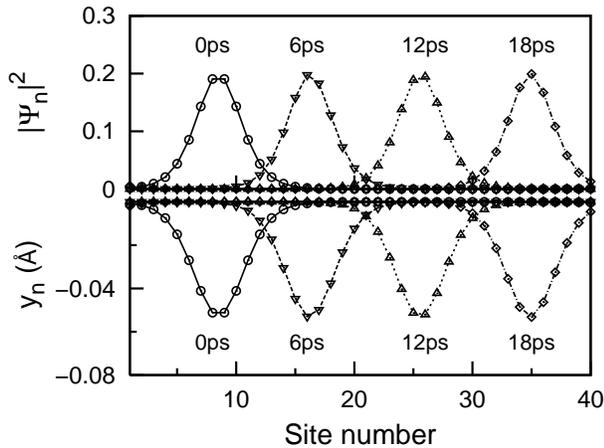}
\caption{Polaron dynamics: the charge density $|\Psi_n|^2$ and 
the lattice displacement $y_n$ propagating through poly-DNA for 
$E$=0.016~mV/\AA ($\chi$=0.6~eV/\AA, $V$=0.1~eV, $m$=300~amu).} 
\label{fig:moving}
\end{figure}

In our simulations, a polaron is initially created by solving 
the system of nonlinear equations ~(\ref{eq:schrod}) and 
~(\ref{eq:newton}) in a stationary situation 
(i.e., for ${d^2y_n}/{dt^2} = {dy_n}/{dt} = 0$ and 
$i\hbar\,{d\Psi_n}/{dt}=$ constant). Initial estimates 
for $y_n$ and $\Psi_n$ are chosen to be nonzero only for ten 
consecutive base pairs. Due to the uniformity and stability of the 
system, the solution is largely insensitive to the initial estimates. 
The resulting initial solution can be seen in Fig.~\ref{fig:moving} 
for $t=0$ ps. The lattice displacement $y_n$ and the charge density 
$|\Psi_n|^2$ coincide and have the same overall shape. Together they 
comprise a polaron that, for the parameters chosen here, 
is spread over approximately thirteen lattice sites but has 
a definite peak in the center. In the polaron's vicinity, the 
displacement $y_n$, and therefore the distortion energy $\chi y_n$, 
are negative, creating a quantum well in the otherwise 
uniform energy profile of the poly-DNA. The wave function $\Psi_n$ is 
consequently localized within this well and the polaron is stable 
against small perturbations. Moreover, if no electric field is applied 
($E$=0), the polaron remains completely stationary because the 
initial solution corresponds to a stable equilibrium.

In an electric field the positively charged polaron 
moves in the direction of the field (Fig.~\ref{fig:moving}) 
for a constant electric field of $E$=0.016~mV/\AA. 
Clearly, as the polaron migrates both the charge density 
and lattice displacement move in unison, i.e., the wave function 
and its formed quantum well travel together. Because of the 
stabilizing effect of the lattice distortion, the polaron retains the 
same overall shape as it migrates; however, our calculations indicate 
that in larger electric fields, the polaron becomes slightly more 
localized over time.

In the following, we study polaron shape and propagation in poly-DNA 
under the influence of an electric field. 
We focus specifically on the effect of three physical parameters, 
$\chi$, $V$, and $m$. The value of the charge-vibrational 
coupling constant $\chi$ determines the decrease in on-site energy 
in the charged-state geometry, and has been estimated theoretically 
to be in the range of 0.3 - 1.5~eV \cite{my_paper2}. Its value 
predominantly depends on the nature of the state geometry and its 
extension, which can be influenced by the structural parameters of 
DNA and the solvent environment as well. The charge in DNA  
can be spread in two direction: in parallel to the propagation pathway 
-- longitudinal direction and perpendicular -- transverse direction. 
The spreading of the charge in the longitudinal direction significantly 
decreases $\chi$ \cite{my_paper1}. 
For the transverse case, a charge can occupy a single 
purine base that suggests a larger value of $\chi$ than does a charge 
that is partially delocalized over a base pair \cite{my_paper2}. 
The degree of charge delocalization also governs the value of 
the polaron mass $m$. If the charge is localized on a single base 
then $m\approx$150~amu, and this value increases with extension of the 
geometry of the state in the transverse direction. The value of 
the charge transfer integral $V$ depends strongly on the configuration 
of the DNA geometry, and can be influenced by the solvent 
environment as well \cite{lewis1}. Theoretical estimations place the 
value of $V$ in the range of 0.05 $-$ 0.3~eV \cite{siebelas,my_paper1}, 
while experiments indicate the value of $V$ to be $\ll$0.01 ~eV 
\cite{lewis1,barton}. Throughout this paper, we use the values 
$\chi$=0.6~eV/\AA, $V$=0.1~eV, and $m$=300~amu as reference points, 
and consider the effects of deviating from these values as we study 
polaron shape and polaron response to applied electric fields.

\begin{figure}
\includegraphics[width=0.44\textwidth]{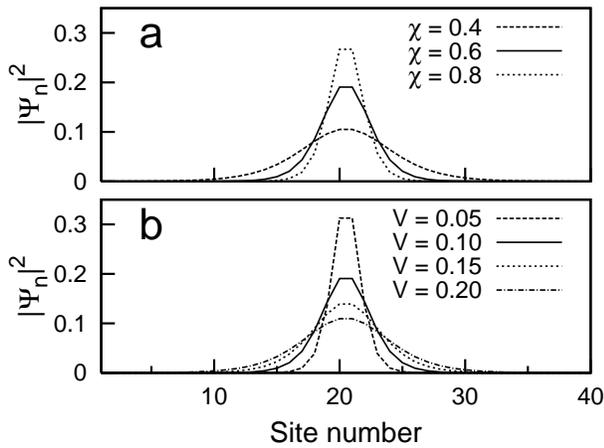} 
\caption{Polaron shape in poly-DNA for varying values of (a) $\chi$ 
and (b) V. In (a), $V$=0.1~eV, $m$=300~amu, and $\chi$ is given in 
units of eV/\AA. In (b), $\chi$=0.6~eV/\AA, $m$=300~amu, and $V$ 
is given in units of eV.} 
\label{fig:shapes}
\end{figure}

The uniform system used in our simulations is ideal for studying 
the effect of model parameters on polaron shape since, as mentioned 
above, the polaron remains stationary in the absence of an 
electric field. Figure~\ref{fig:shapes}(a) shows the effect of 
changing the coupling constant $\chi$ under these circumstances. 
As is evident in the figure, increasing $\chi$ results in greater 
polaron localization: the polaron occupies fewer lattice sites, 
with a correspondingly larger charge density and lattice distortion 
at the center of the polaron. Conversely, decreasing $\chi$ has 
the opposite effect. Figure~\ref{fig:shapes}(b) shows the shape 
of the polaron at varying values of the charge transfer integral 
$V$. The effect is opposite to that observed when adjusting $\chi$: 
increasing (decreasing) $V$ causes a decrease (increase) in polaron 
localization. This is expected since the larger the charge 
transfer integral, the more the polaron will spread out to neighboring 
sites. In accordance with Eq.~(\ref{eq:newton}), the value of $m$ has 
no effect on the polaron's shape: in the stationary state the time 
derivative $m\gamma\,{dy_n}/{dt}$ in Eq.~(\ref{eq:newton}) equals zero.

After the polaron is initially created, a constant uniform electric 
field $E$ is applied and we study the resulting polaron motion in time. 
Simulations show that there exists a maximum electric field $E_{max}$ 
that the polaron can tolerate; for $E>E_{max}$, the lattice displacement 
and charge density become unsynchronized and irregular. The value of 
$E_{max}$ depends on the system's parameters and in all cases studied, 
the more delocalized the polaron the smaller the value 
of $E_{max}$. As such, increasing $\chi$ or decreasing $V$ leads to a 
polaron capable of tolerating larger fields, whereas $E_{max}$ is 
mostly independent of $m$. A highly localized polaron causes a large 
localized lattice distortion, which effectively acts as a quantum well. 
By decreasing the energy in its vicinity, a more localized polaron is 
therefore more stable, and consequently, it is reasonable that the 
polaron would be able to remain well-formed in the presence of 
stronger electric fields.

The application of an electric field effects polaron migration 
(Fig.~\ref{fig:moving}). Moreover, provided that $E~<~E_{max}$, 
the polaron remains well-formed and the motion is continuous. 
Further, in the presence of a field the polaron moves at a 
constant velocity. This result can be explained as follows: (i) our 
system represents poly-DNA under periodic boundary conditions, so the 
system parameters do not vary with location; (ii) the applied electric 
field is uniform and constant; (iii) the polaron maintains a constant 
shape as it travels; and (iv) the friction term $m\gamma\,{dy_n}/{dt}$ 
prevents the charge from continually accelerating. These factors ensure 
that after a brief period of acceleration from the initial stationary 
state, the polaron's velocity does not vary. 

\begin{figure}
\includegraphics[width=0.44\textwidth]{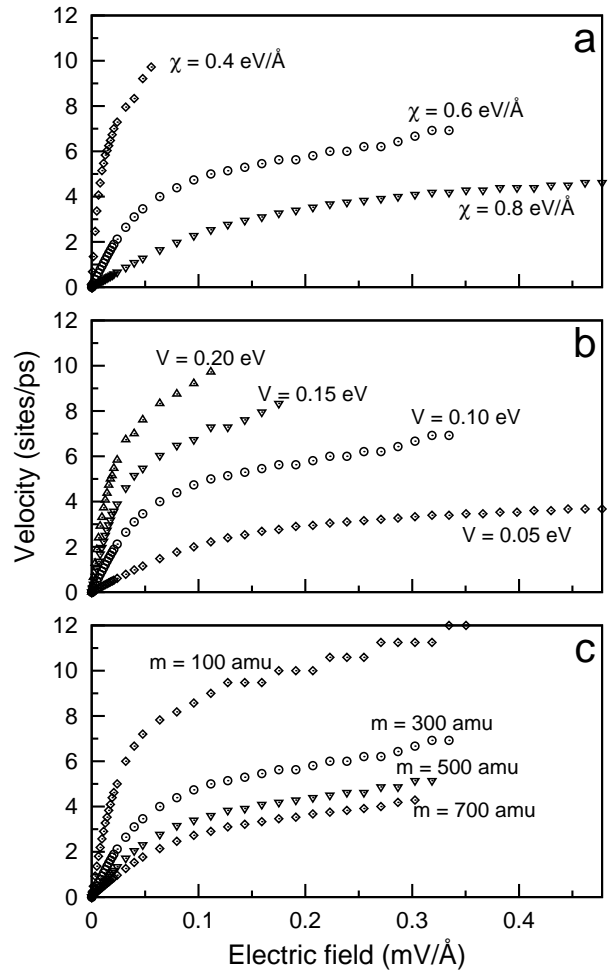}
\caption{Propagation velocity of a polaron through poly-DNA as a 
function of applied electric field for varying values of (a) $\chi$, 
(b) $V$, and (c) $m$. Points corresponding to $E~>~E_{max}$ are absent as 
they do not represent accurate values. Unless specified otherwise in 
the figure, $\chi$=0.6~eV/\AA, $V$=0.1~eV, and $m$=300~amu.}
\label{fig:velocities}
\end{figure}

The velocity does depend, however, on the system parameters and 
on the applied electric field strength. Figures ~\ref{fig:velocities}(a), 
~\ref{fig:velocities}(b), and ~\ref{fig:velocities}(c) show how the 
velocity of the polaron varies with $E$ for different values of $\chi$, 
$V$, and $m$ respectively. These graphs all display a similar behavior, 
but the magnitudes and slopes in the plots are very sensitive 
to the values of $\chi$, $V$, and $m$. In particular, an increase 
in the velocity $-$ and therefore conductance $-$ can be achieved 
by decreasing the coupling constant $\chi$, by increasing the 
charge transfer integral $V$, or by decreasing the polaron mass $m$. 
Analysis of both $\chi$ and $V$ suggest that an increase in polaron 
localization produces a decrease in velocity. It should be noted, 
however, that this correlation need not be a general mathematical rule; 
for example, by changing $m$ it is possible to change the polaron's 
velocity without affecting its localization at all. Other model 
parameters, such as the friction constant $\gamma$, also affect 
the velocity, but do not change the qualitative behavior observed 
in Fig.~\ref{fig:velocities}.

\begin{figure}
\includegraphics[width=0.44\textwidth]{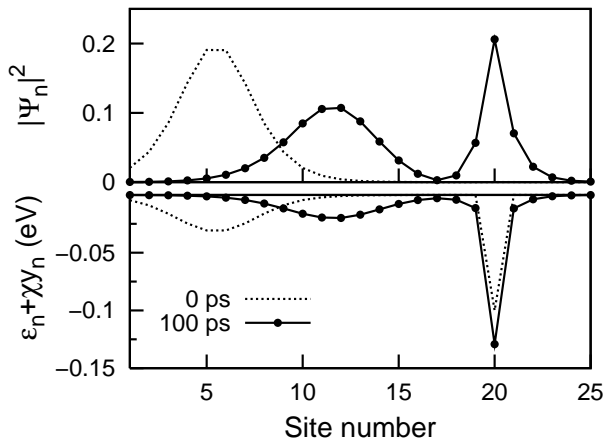}
\caption{Effect of an irregularity: the charge density $|\Psi_n|^2$ and 
the site energy $\chi y_n+\epsilon_n$ propagating through poly-DNA with 
a disorder-induced well. Here $\epsilon_{20}$=0.1~eV and $\epsilon_n$=0 
for n$\neq$20. At $t$=0~ps, $\chi y_{20}$=0~eV; at $t$=100~ps, $\chi 
y_{20}$=0.029~eV.}
\label{fig:blip}
\end{figure}

Structural disorder in the poly-DNA can drastically alter 
the polaron propagation. An irregularity at a base site $k$ can 
produce a quantum barrier or well, which we represent with a non-zero 
value of $\epsilon_k$. If $\epsilon_k>$~0 (a barrier), the polaron is 
usually unable to traverse site $k$ for any $E<E_{max}$, instead coming 
to a stop just before the irregularity. Only for very low barriers 
($\epsilon_k\approx$0.01~eV) in conjunction with the polaron size 
and a large electric field ($E\approx$0.1~mV/\AA) can the large 
polaron migrate through the barrier. The situation for quantum wells 
($\epsilon_k<$0) is very different. For very shallow wells 
($\epsilon_k\gtrsim$-0.05~eV), the polaron completely enters the well 
for any $E$, although a large $E$ can dislodge the polaron if the 
well is extremely shallow. For deeper wells, the polaron sometimes 
splits into two, with one polaron halting prior to the well and second 
entering the well (Fig.~\ref{fig:blip}). In these cases, 
the initial polaron stops before the irregularity, but then part of 
it tunnels from the polaron-induced well ($\chi y_n$) into the 
disorder-induced well ($\epsilon_k$). It should be emphasized that 
the above analysis applies only to poly-DNA. In less uniform situations, 
the system is generally unstable, and the polaron often tunnels 
through barriers to reach distant wells \cite{my_paper2}. 

In conclusion, the velocity of polaron propagation and hence 
the DNA conductance is mostly determined by the polaron size in 
transverse and longitudinal directions. A decrease of $\chi$ induced 
by partial delocalization of the charge from the purine base to the 
whole base pair \cite{my_paper1} provides the extension of 
polaron size in the longitudinal direction and significantly increases 
the DNA conductance. For example, a decrease of $\chi$ by 0.2 eV can 
increase the conductance by $\sim$ 6 -- 10 times depending on the electric 
field value. The polaron can be destroyed at high electric fields,  
which then changes the mechanism of charge transfer in the DNA molecule 
and therefore, will cause a discontinuity at the conductance characteristics.
A large polaron has higher velocity but is less tolerant to the 
electric field. The structural disorders, which form a barrier $\lesssim$ 
0.05eV (well) on the polaron pathway, depending on energetic conditions
and electric field magnitude, can stop (trap) the polaron and 
cause a fast decrease of DNA conductance resulting in an insulating 
behavior. The influence of disorder on the DNA conductance decreases 
with increasing temperature.

\section*{Acknowledgments}
The work has been supported by the Canada Research Chair Program
and the NSERC Discovery Grant.


\begin{thebibliography}{99}
\bibitem[\ddag]{byline} Electronic mail: tapash@physics.umanitoba.ca
\bibitem{review} {\it Charge Migration in DNA: Perspectives from Physics,
Chemistry, and Biology}, edited by T. Chakraborty (Springer, New York,
2007); {\it Long-range charge transfer in DNA}, edited by G.B. Schuster 
(Springer-Verlag, Heidelberg, New York, 2004).
\bibitem{jortner}
J. Jortner, M. Bixon, T. Langenbacher, and M.E. Michel-Beyerle, 
Proc. Natl. Acad. Sci. (USA) {\bf 95}, 12759, (1998). 
\bibitem{berlin}
Y.A. Berlin, A.L. Burin, and M.A. Ranter, J. Phys. Chem. A 
{\bf 104}, 443 (2000).
\bibitem{yoo}
K.-H. Yoo, et al., Phys. Rev. Lett. {\bf 87} 198102 (2001).
\bibitem{kasumov}
A.Y. Kasumov, et al. Science {\bf 291} 280 (2001).
\bibitem{kawai}
M. Taniguchi, and T. Kawai, Physica E {\bf 33}, 1 (2006).
\bibitem{porath}
D. Porath, A. Bezryadin, S. deVries, C. Dekker, Nature 
{\bf 403}, 635 (2000).
\bibitem{conw}
E. Conwell, Top. Curr. Chem. {\it 237}, 73 (2004).
\bibitem{bishop1}
P. Maniadis, et al. Phys. Rev. E {\bf 72}, 021912 (2005).
\bibitem{lewis}
F.D. Lewis, et al. Angew. Chem. Int. Ed. {\bf 45}, 7982 (2006).
\bibitem{giese}
B. Giese, et al. Nature {\bf 412}, 318 (2001).
\bibitem{su}
W.P. Su, J.R. Schrieffer, A.J. Heeger, Phys. Rev. B. {\bf 22}, 2099 (1980).
\bibitem{Ono}
Y. Ono, and A. Terai, J. Phys. Soc. Jap. {\bf 59}, 2893 (1990).
\bibitem{my_paper1}
J. Berashevich, and T. Chakraborty, Chem. Phys. Lett. (2007) in press.
\bibitem{my_paper2}
J. Berashevich, and T. Chakraborty, cond-mat/0709.0954.
\bibitem{siebelas}
K. Senthilkumar {\it et al.}, J. Am. Chem. Soc. {\bf 127}, 14894 (2005).
\bibitem{lewis1}
F.D. Lewis, et al. J. Am. Chem. Soc. {\bf 122} 12346 (2000).
\bibitem{barton}
M. A. O'Neill, et al. Angew. Chem. Int. Ed. {\bf 42}, 5896 (2003).
\end{thebibliography}
\end{document}